\documentclass{PoS}

\newcommand{\sign}{\mbox{sign}}

\title{Improvement of algorithms for dynamical overlap fermions}

\ShortTitle{Improvement of algorithms for dynamical overlap fermions}

\author{JLQCD Collaboration: 
  Hideo~Matsufuru$^{a}$\thanks{Speaker (Email: hideo.matsufuru@kek.jp).}, 
  Hidenori~Fukaya,$^b$
  Shoji~Hashimoto,$^{ac}$
  Kazuyuki~Kanaya,$^d$
  Takashi~Kaneko,$^{ac}$
  Kenji~Ogawa,$^e$
  Masataka~Okamoto,$^a$
  Tetsuya~Onogi,$^f$
  Norikazu~Yamada.$^{ac}$\\
  \llap{$^a$}High Energy Accelerator Research Organization (KEK),
             Tsukuba 305-0801, Japan.\\
  \llap{$^b$}Theoretical Physics Laboratory, RIKEN, Wako 351-0198, Japan.\\
  \llap{$^c$}School of High Energy Accelerator Science,
             The Graduate University for Advanced Studies (Sokendai),
             Tsukuba 305-0801, Japan.\\
  \llap{$^d$} Graduate School of Pure and Applied Sciences,
           University of Tsukuba, Tsukuba 305-8571, Japan\\
  \llap{$^e$}Department of Physics, National Taiwan University,
      Taipei 10617, Taiwan.\\
  \llap{$^f$}Yukawa Institute for Theoretical Physics, Kyoto University,
             Kyoto 606-8502, Japan.}

\abstract{
We investigate the algorithms for dynamical overlap fermions
aiming at improving the performance for large-scale simulations. 
We look for the best combination of Hybrid Monte Carlo options 
and iterative quark solvers with respect to the numerical costs. 
Our main target is a $N_f=2$ simulation with overlap fermion 
on a $16^3\times 32$ lattice at lattice spacing around 0.12~fm.}

\FullConference{XXIVth International Symposium on Lattice Field Theory\\
                July 23-28, 2006\\
                Tucson, Arizona, USA}

\begin{document}


\section{Introduction}

The JLQCD Collaboration is performing dynamical QCD simulations
with the overlap fermions, as a new project started in 2006
\cite{Kaneko,Hashimoto,Yamada,Fukaya}.
At present, our main run is generating lattices of size
$16^3 \times 32$, $a\simeq 0.12$~fm, with
two flavors of sea quarks whose smallest mass $\simeq m_s/6$.
The topological sector is fixed by a pair of extra Wilson fermions.
This considerably improves the efficiency of the HMC algorithm,
while the effect of fixing the topological charge $Q$
should be examined by measuring
on configurations with different values of $Q$.
Numerical simulations at different values of $Q$, as well as with
larger lattices and with $2+1$ flavors, are also in preparation.

These studies are being carried out on a new supercomputer
system at KEK,
which is in service since March 2006 \cite{KEKSC}.
The system has two computational servers:
Hitachi SR11000 model K1 (peak performance 2.15TFlops),
and IBM System Blue Gene Solution (57.3TFlops).
The latter system has 10 racks, each composed of 1024
nodes (2048 processor cores).
For the Wilson fermion solver, with data on cache,
the Blue Gene system achieves about 29\%
of the peak performance on a half-rack system.%
\footnote{
We thank J.~Doi and H.~Samukawa of IBM Japan for tuning the
Wilson kernel on the Blue Gene system.}

Even on these powerful machines, the dynamical overlap
simulation is quite challenging.
The simulation is performed with the Hybrid Monte Carlo
(HMC) algorithm.
In this paper, we describe our attempt to speed-up the
simulation by improving the HMC algorithms.


\section{Action}

The effective action we treat in the HMC simulation has a form,
\begin{equation}
S = S_G + S_F + S_E .
\label{eq:action}
\end{equation}
$S_G$ is the gauge field part, for which we adopt the
renormalization group improved (Iwasaki) action, while
in an exploratory stage of this work
a modified plaquette gauge action, which is designed to suppress
dislocations, was also investigated \cite{Yamada}.

As the quark action, we use the $N_f=2$ overlap fermion action.
The overlap-Dirac operators is represented as
\begin{equation}
D(m) = \left(m_0 + \frac{m}{2} \right)
  + \left( m_0 - \frac{m}{2}\right) \gamma_5 \, \mbox{\rm sign}(H_W) ,
\label{eq:overlap_opr}
\end{equation}
where $H_W=\gamma_5 D_W$, $D_W$ is the Wilson-Dirac operator
with a large negative mass $-m_0$.
The sign function in Eq.~(\ref{eq:overlap_opr})
is approximated by a partial fraction form
\cite{vandenEshof:2002ms,Chiu:2002eh}:
\begin{equation}
\mbox{\rm sign} (H_W) = 
\frac{H_W}{\sqrt{H_W^2}} \simeq H_W
  \left( p_0  + \sum_{l=1}^N \frac{p_l}{H_W^2 + q_l} \right) .
\label{eq:Zolotarev}
\end{equation}
\noindent
The $N$ inversions $(H_W^2+q_l)^{-1}$ are calculated at the
same time using the multi-mass conjugate gradient method.

We utilize the mass preconditioning \cite{Hasenbusch:2001ne},
{\it i.e.} introducing a preconditioning term with heavier quark
mass $m'$ than that of the dynamical quark. 
The fermion action becomes $S_F = S_{PF1} + S_{PF2}$,
\begin{equation}
 S_{PF1} = \phi_1^\dag [D(m')^\dag D(m')]^{-1} \phi_1,
 \hspace{0.3cm}
 S_{PF2} = \phi_2^\dag \left\{ D(m')
          [D(m)^\dag D(m)]^{-1}  D(m')^\dag \right\} \phi_2 ,
\end{equation}
where $S_{PF1}$ is the preconditioner and $S_{PF2}$ is
the preconditioned dynamical quark term with corresponding
pseudofermion fields $\phi_1$ and $\phi_2$, respectively.

$S_E$ is the extra Wilson fermion term defined as
\begin{equation}
\det \left( \frac{H_W^2}{H_W^2+\mu^2} \right)
     = \int {\cal D}\chi^\dag {\cal D}\chi \exp[- S_E] ,
\label{eq:extraWilson}
\end{equation}
\begin{equation}
 S_E = \chi^\dag \left[
  (D_W+i\gamma_5 \mu) (D_W^\dag D_W)^{-1} (D_W+i\gamma_5 \mu)^\dag
 \right] \chi ,
\end{equation}
where the denominator of Eq.~(\ref{eq:extraWilson}) amounts to
two flavors of heavy ghosts with a twisted mass $\mu$.
This term suppresses near-zero modes of $H_W$,
while keeping the effects on higher modes minimal
\cite{Hashimoto,Vranas:1999rz,Fukaya:2006ca,Fukaya:2006vs}.
The newly introduced fields have mass of the order of lattice
cutoff and therefore irrelevant in the continuum limit.

\begin{figure}[tb]
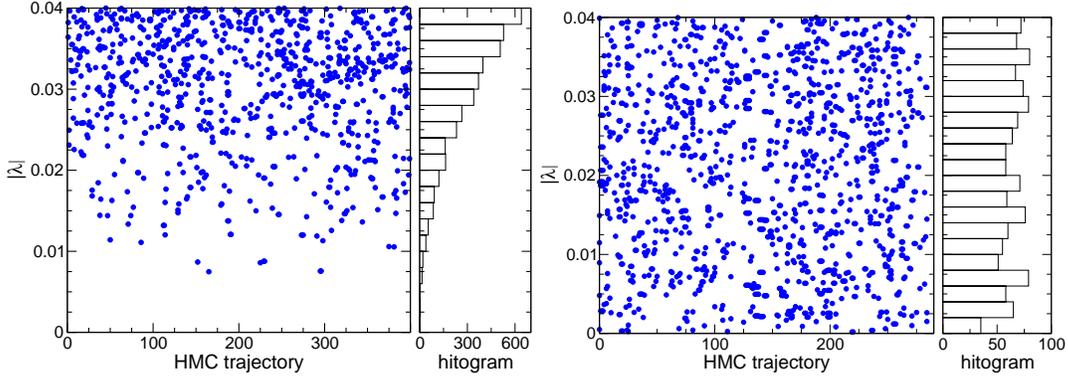

\center{
\includegraphics[clip=true,width=7.cm]{Figs/fig1.eps}
\includegraphics[clip=true,width=7.cm]{Figs/fig2.eps}}
\vspace{-0mm}
\caption{
The low-lying modes of $H_W$ in the case
with $S_E$ ($\mu=0.2$, left panel) and without
$S_E$ (right panel) at
$a\simeq 0.125$~fm and $m\simeq 2m_s$.
}
\label{fig:lowmodes1}
\vspace{-0mm}
\end{figure}

The quark action becomes singular when
$H_W$ has a zero eigenvalue.
This causes discontinuity in the conjugate momenta when
$\lambda_{min}$ changes the sign during the molecular dynamics
evolution.
While this problem can be circumvented by the so-called
reflection/refraction prescription
\cite{Fodor:2003bh}, it requires monitoring of 
the near-zero eigenvalues and additional inversions of overlap
operator, which largely increase the numerical cost.
In our case, however, extra Wilson fermions prevent the near-zero mode
from approaching zero, and hence these operations can be skipped.

\begin{figure}[tb]
\center{
\includegraphics[clip=true,width=7.cm]{Figs/mdeigen_b2.35_m110.eps}
\includegraphics[clip=true,width=7.cm]{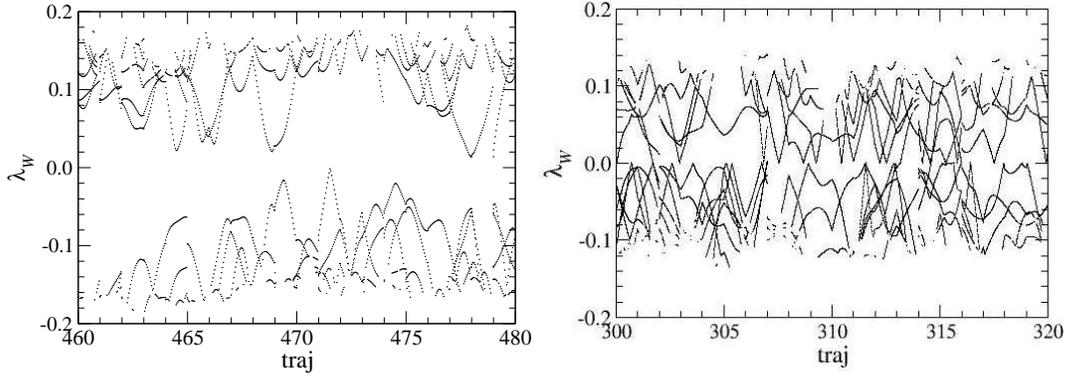}}
\vspace{-0mm}
\caption{
Evolution of $\lambda_{min}$ in the molecular dynamics at
$a\simeq 0.12$~fm and $m\simeq 2m_s$.
The left and right panels show the cases with and without
$S_E$, respectively.
The left panel shows an event that as if a reflection occurs, while
is actually not the case.
}
\label{fig:lowmodes2}
\vspace{-0mm}
\end{figure}

How the extra Wilson fermions work is depicted in
Figure~\ref{fig:lowmodes1}.
The figure compares appearance of low-lying eigenmodes
during the HMC runs with and without $S_E$ at
$a\simeq 0.125$~fm and $m\simeq m_s$.
It is clear that $S_E$ suppresses the spectral density around
$\lambda=0$.
The same feature is found in the molecular dynamics history
of the lowest mode displayed in Figure~\ref{fig:lowmodes2}.
With $S_E$, no reflection nor refraction occurs, contrary to
the case without $S_E$.
One can therefore switch off the monitoring of $\lambda$ in the
case with $S_E$.
Even when $\lambda=0$ occurs due to a finite molecular dynamics
step size, it is signaled by large $\Delta H$
and thus rejected in the Metropolis test.

\section{Solver algorithms}
 \label{sec:solver}

Since the inversion of the overlap-Dirac operator is
the most time consuming part of the HMC simulation,
improvement of the solver algorithm is crucial.
We compare two methods: the nested CG with relaxed precision
of the inner CG loop, and the 5-dimensional CG algorithm.

The overlap operator requires computation of the partial fraction
terms in Eq.~(\ref{eq:Zolotarev}).
Therefore, the CG method to invert the overlap operator has
a nested structure; the inner loop to calculate
$(H_W^2+q_l)^{-1}$, and the outer loop to operate $D(m)$.
For the inner loop, the multi-shift CG method
is used to solve $(H_W^2+q_l)^{-1}$ for all $l$ simultaneously.
The precision of the approximation Eq.~(\ref{eq:Zolotarev})
is determined by the degree $N$ and the condition number
$\lambda_{max}/\lambda_{min}$.
For a smaller $|\lambda_{min}|$, a larger $N$ is needed to keep
the precision;
{\it e.g.} $N$=10 corresponds to $O(10^{-7})$ accuracy
for $|\lambda_{min}|$=0.05 and $O(10^{-5})$ for 0.01.
The multi-shift CG method has an advantage
that the cost is almost independent of $N$.
Instead of extending the window $[|\lambda_{min}|,|\lambda_{max}|]$
for small $|\lambda_{min}|$, we may project out the low-lying modes
explicitly and add back with the eigenvalue $\sign(\lambda)$.
In this way we may fix the lower limit of the approximation to
some threshold $\lambda_{thrs}$, below which the eigenmodes are
treated exactly.

The relaxed CG method is an improvement of the nested CG method.
It changes the precision of the inner loop adaptively as the
outer loop iteration proceeds \cite{Cundy:2004pz}.
As we will see, the relaxed CG is about twice faster than
the original CG.

An alternative solver is the 5-dimensional CG method
\cite{Edwards:2005an}.
Let us consider the following form of a 5D matrix (an explicit
example for the $N=2$ case):
\begin{equation}
 M_5 =
  \left( \begin{array}{cccc|c}
               H_W        & -\sqrt{q_2} & & & 0  \\
            -\sqrt{q_2} & -H_W          & & & \sqrt{p_2} \\
               & &    H_W        & -\sqrt{q_1} & 0  \\
               & & -\sqrt{q_1} & -H_W          & \sqrt{p_1} \\
    \hline
             0 &\sqrt{p_2} & 0 & \sqrt{p_1} & R \gamma_5 + p_0 H_W \\
         \end{array}   \right)
 = 
  \left( \begin{array}{c|c}
               A   & B  \\
    \hline
               C   & D  \\
         \end{array}   \right) .
\end{equation}
Each component represents the usual 4D matrix.
By the Schur decomposition,
\begin{equation}
 M_5 = \tilde{L} \tilde{D} \tilde{U} =
  \left( \begin{array}{cc}
             1 & 0 \\
             CA^{-1} & 1 \\
         \end{array}   \right)
  \left( \begin{array}{cc}
             A & 0 \\
             0 & S \\
         \end{array}   \right)
  \left( \begin{array}{cc}
             1 & A^{-1}B \\
             0 & 1 \\
         \end{array}   \right) ,
\end{equation}
where
\begin{equation}
 S = D - CA^{-1}B 
   = R \gamma_5 + p_0 H_W + H_W \sum_i \frac{p_i}{H_W^2+ q_i}
\end{equation}
expresses the partial fraction approximation of $D(m)$.
Therefore, by solving
\begin{equation}
 M_5
  \left( \begin{array}{c}
             \phi   \\
             \psi_4 \\
         \end{array}   \right)
= \left( \begin{array}{c}
               0   \\
             \chi_4 \\
         \end{array}   \right) ,
\label{eq:5dim_lineq}
\end{equation}
$\psi_4 = S^{-1} \chi_4$ is determined.
A preconditioning is applied by multiplying
the inverse of $M_5^{(0)}=M_5[U=0]$, which is easily inverted
by forward and backward substitutions.
The even-odd preconditioning is also applicable, and according
to our performance comparison, this is the best solution for
the 5D solver.
Since the size of $M_5$ grows as $N$, the numerical cost
increases linearly in $N$.
A disadvantage is that the subtraction of low-modes of $H_W$
is not applicable when the even-odd decomposition is used.
This causes a difficulty when $\lambda_{min}$ becomes too small
to be approximated without the projection.
To apply the 5D solver, one needs to determine the lowest
boundary $V_{min}$ above which the partial fraction approximation
is valid.

\begin{figure}[tb]
\vspace{-0.5cm}
\center{
\includegraphics[width=8.0cm]{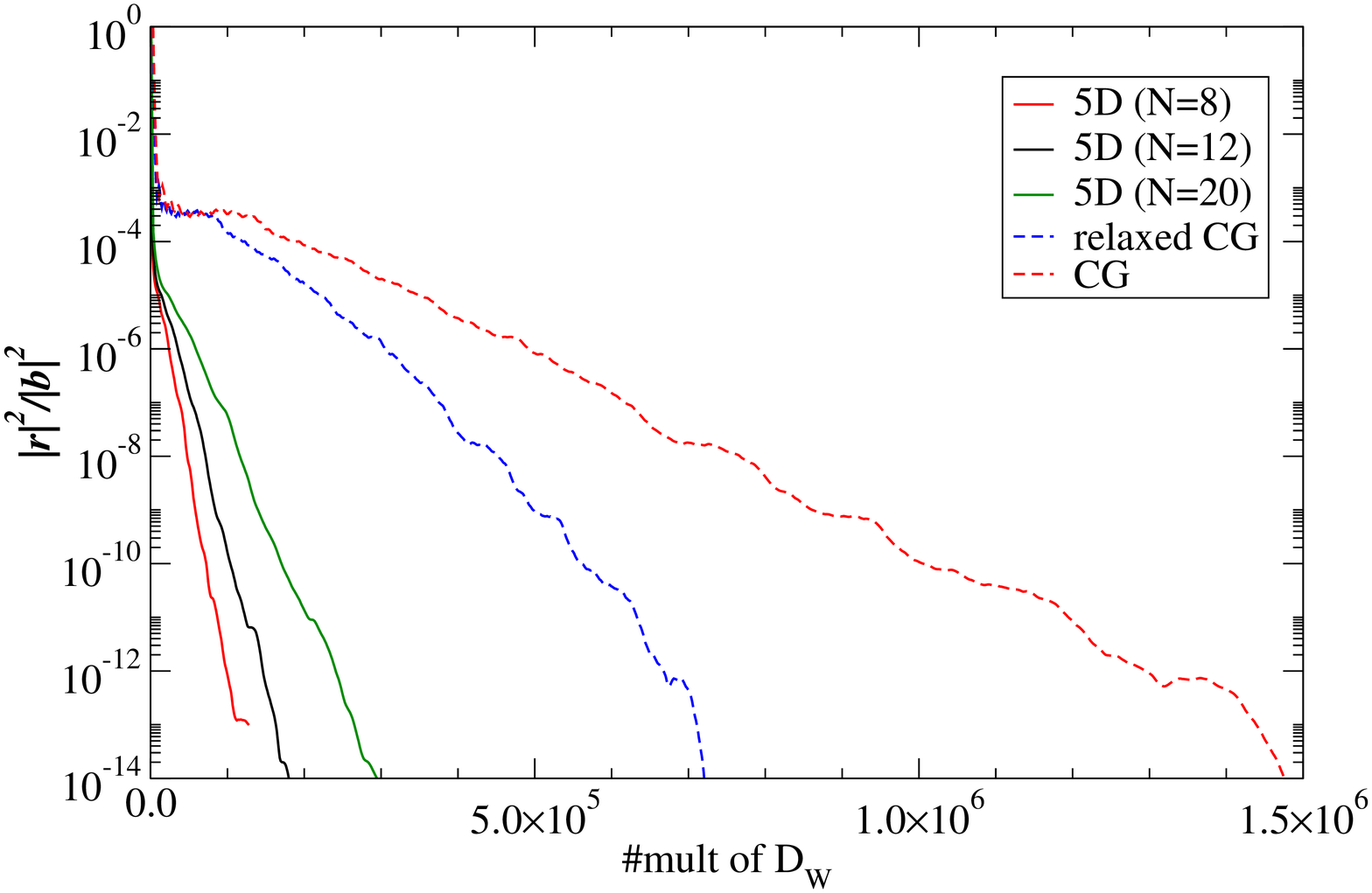}
\includegraphics[width=7.0cm]{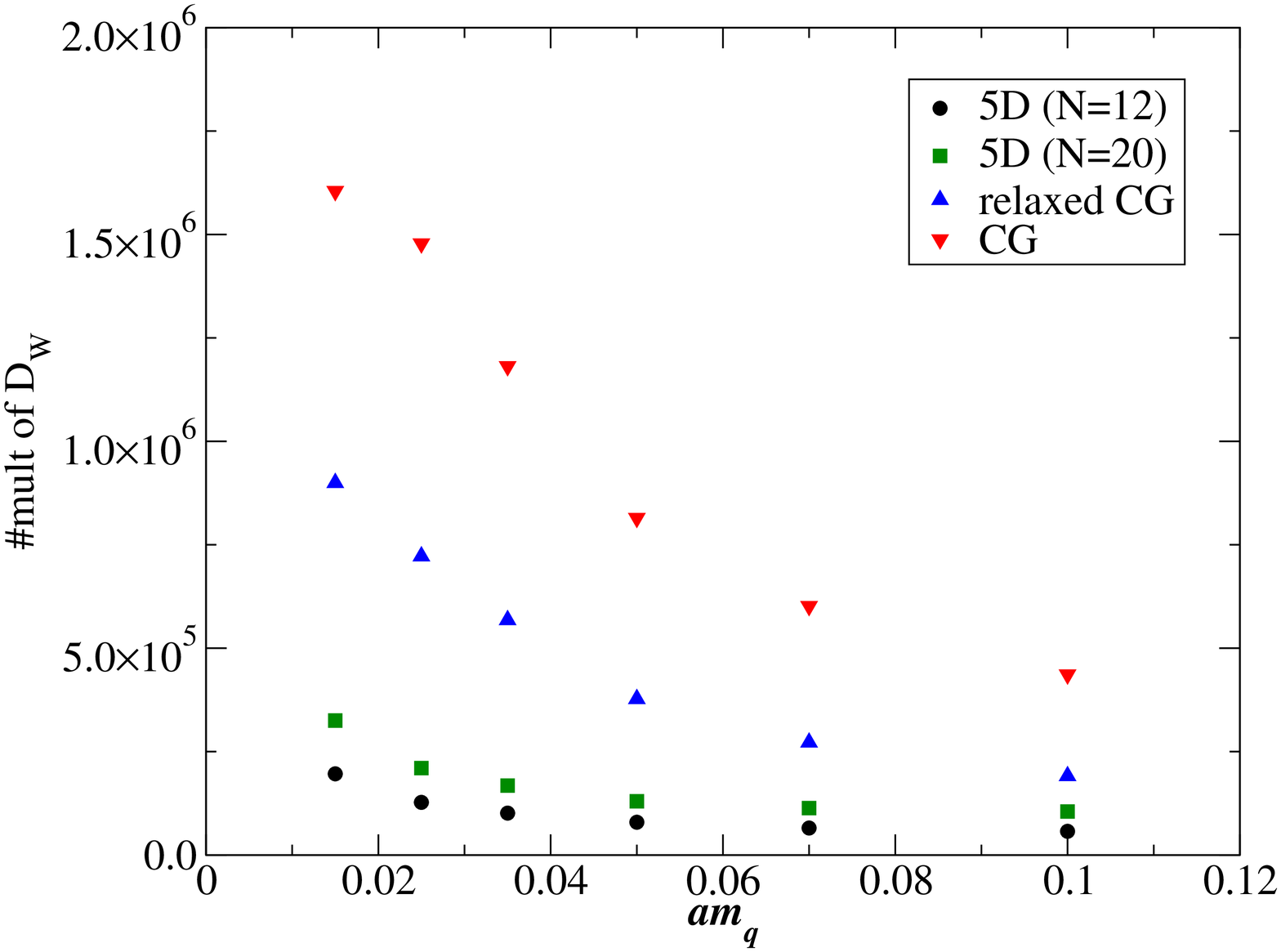}}
\vspace{-0.8cm}
\caption{
Comparison of the solver algorithms on a single configuration.
The left panel shows convergence of the residue as the number
of $D_W$ multiplication.
The right panel shows quark mass dependence of the number
of $D_W$ multiplication needed for $|r|^2/|b|^2<10^{-14}$.
}
\label{fig:solver}
\vspace{-0mm}
\end{figure}

Figure~\ref{fig:solver} shows the comparison of the solver algorithms
at $a\simeq 0.12$~fm, $m\simeq 0.4 m_s$, on a single
configuration.
The figure shows that the relaxed CG is factor of 2 faster than the
standard CG method.
The 5D solver is even faster by another factor of 2--3
than the relaxed CG for $N=20$.
This conclusion is independent of the quark mass, as displayed in
the right panel of Fig.~\ref{fig:solver}.
Therefore, if near-zero eigenvalues do not appear,
as in the present case, the 5D solver is the fastest.


\section{HMC algorithm}
 \label{sec:HMC}

\paragraph{Multi-time step.}

Magnitude of the forces corresponding to the terms,
$S_G$, $S_{PF1}$, $S_{PF2}$, and $S_E$ 
has a hierarchical structure.
In particular the gauge part has the largest contribution to
the evolution of the conjugate momenta, while the cost to
compute it is negligible compared to the fermionic part.
The size of the force for $S_{PF2}$ is smaller compared to
that of $S_{PF1}$.
The multi-time step \cite{Sexton:1992nu}
makes use of this hierarchy by adopting different time
steps for these terms in the molecular dynamical evolution.

\begin{figure}[tb]
\center{
\includegraphics[clip=true,width=7.cm]{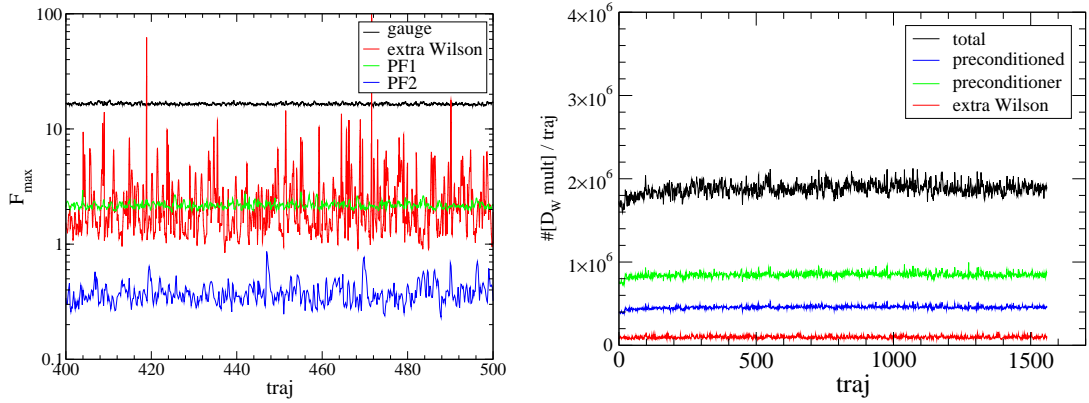}
\includegraphics[clip=true,width=7.3cm]{Figs/mdcost.eps}}
\vspace{-0mm}
\caption{
The maximum values of the forces (left panel) and
the costs of the forces (right panel) monitored in HMC
at $a\simeq 0.12$~fm and $m\simeq 2m_s$.
}
\label{fig:force}
\vspace{-0mm}
\end{figure}

The forces are compared in Fig.~\ref{fig:force}.
This result suggest to chose the step sizes as
\begin{eqnarray}
\Delta\tau_{(PF2)} \gg \Delta\tau_{(PF1)} \gg
 \Delta\tau_{(G)} = \Delta\tau_{(E)}.
\end{eqnarray}
While the size of the force for $S_E$ is as small as the fermionic
part, $\Delta\tau_{(E)}$ is set to be the same as the gauge part,
to ensure the disappearance of the near-zero modes,
because the fluctuation of the $S_E$ force is large.
The cost to determine the force of the extra Wilson fermions
is negligible compared to the overlap fermion part.
The ratio of the step sizes are determined by monitoring the size of
the forces.
For example, $\Delta\tau_{(PF2)}/\Delta\tau_{(PF1)} = 5$
and $\Delta\tau_{(PF1)}/\Delta\tau_{(G,E)} = 6$ are a reasonable
choice for the displayed case.

\paragraph{Noisy Metropolis test.}

Considering the performance of the solvers in Sec.~\ref{sec:solver},
the 5D CG method is preferable, with small number of poles $N$
if possible.
As for the preconditioner, we can choose relatively small $N$,
since the contributions to the dynamics cancel in $S_{PF1}$
and $S_{PF2}$.
For $S_{PF2}$, one can also choose an $N$ with a less precise
approximation by making use of the noisy Metropolis algorithm
\cite{Kennedy:1985pg}, which is prescribed as follows.
At the end of a molecular dynamics evolution, 
after the usual Metropolis test,
we accept $U_{new}$ with a probability $P = \min\{1,e^{-dS}\}$, where
\begin{eqnarray}
 dS = \left|\, W^{-1}[U_{new}] \, W[U_{old}]\, \xi \, \right|^2 - |\xi|^2 ,
\end{eqnarray}
$W = D(m)/D'(m)$, with $D'$ a less accurate overlap operator used
in HMC, and $D$ the accurate overlap operator,
$U_{old}$ is the initial gauge field, and $\xi$ is a random Gaussian
noise vector.

\paragraph{Performance.}

Finally, we show the present performance of HMC measured
on the Blue Gene (512-node) at $a\sim 0.12$~fm, $\mu=0.2$,
and a trajectory length $\tau = 0.5$.
The first three lines in Table~\ref{tab:performance} show
the result for the simulation with the 4D (relaxed CG) solver,
with which most of gauge configurations are
generated so far.
No noisy Metropolis test is incorporated.
The last three lines in Table~\ref{tab:performance} show
a preliminary result for the performance with fully improved
algorithm, the less precise 5D solver in molecular dynamics
with the noisy Metropolis test, which achieves about a factor
of 3 acceleration.
Therefore this algorithm is our current best
option, which will be adopted in our productive run in future.

\begin{table}[tb]
\begin{center}
\begin{tabular}{cccccccccc}
\hline
 & $m_{ud}$ & $m'$ &  $N_{\tau(PF2)}$ & 
$\frac{\Delta\tau_{(PF2)}}{\Delta\tau_{(PF1)}}$ &
$\frac{\Delta\tau_{(PF1)}}{\Delta\tau_{(G,E)}}$ &
$N_{PF1}$ & $N_{PF2}$ & $P_{acc}$ & time[min] \\
\hline
Nested CG & 0.015 & 0.2 & 9 & 4 & 5 & 10 & 10 & 0.87 & 112 \\
  (4D)    & 0.025 & 0.2 & 8 & 4 & 5 & 10 & 10 & 0.90 & 94 \\
          & 0.035 & 0.4 & 6 & 5 & 6 & 10 & 10 & 0.74 & 63 \\
\hline
5D solver & 0.035 & 0.4 & 7 & 5 & 6 & 10 & 10 & 0.68 & 22 \\
          & 0.035 & 0.4 & 8 & 5 & 6 & 10 & 10 & 0.80 & 26 \\
          & 0.035 & 0.4 & 8 & 5 & 6 & 6  & 10 & 0.78 & 23 \\
\hline
\end{tabular}
\caption{Performance of HMC on Blue Gene (512-node)
at $a\sim 0.12$~fm, $\mu=0.2$.
Step-1 means the simulation with (4D) nested CG overlap solver,
and Step-2 with the 5D overlap solver corrected by the noisy
Metropolis test.}
\label{tab:performance}
\end{center}
\end{table}

\bigskip

Numerical simulations are performed on Hitachi SR11000 and
IBM Blue Gene at High Energy Accelerator Research
Organization (KEK) under a support of its Large Scale
Simulation Program (No. 06-13).
This work is supported in part by the Grant-in-Aid of the
Ministry of Education 
(No. 13135213, 16740156, 17340066, 17740171, 18034011,
18340075, 18740167).

\end{document}